\documentclass[useAMS,usenatbib]{mn2e}

\usepackage {graphicx}
\def\aj{\,{AJ}}
\def\apj{\,{\rm ApJ}}
\def\apjl{\,{\rm ApJL}}

\def\aap{\,{\rm A\&A}}
\def\nat{\,{\rm nature}}
\def\mnras{\,{\rm MNRAS}}

\title[Evolution of 
the Colour--Radius Relation 
and 
the Morphology--Radius Relation 
in the SDSS]%{Evolution of the Morphology--Clustercentric-Radius Relation	%in the SDSS}
{Evolution of the Colour--Radius Relation  
 and
the Morphology--Radius Relation in the SDSS Galaxy Clusters}
\author[Tomotsugu Goto  et al.]{Tomotsugu Goto$^{1,2}$\thanks{E-mail: 
yohnis@icrr.u-tokyo.ac.jp, tomo@jhu.edu} \thanks{Presnet address: Department of Physics and Astronomy, The Johns Hopkins University, 3400 North Charles Street, Baltimore, MD 21218-2686, USA}, Masafumi Yagi$^{3}$,  Masayuki Tanaka$^{2}$, and
Sadanori Okamura$^{2}$
\\
$^{1}$Institute for Cosmic Ray Research, University of
Tokyo, Kashiwanoha, Kashiwa, Chiba 277-0882, Japan\\
$^{2}$Department of Astronomy, Graduate School of Science, The University of Tokyo,
Hongo 7-3-1, Bunkyo-ku, Tokyo 113-0033, Japan\\
$^{3}$National Astronomical Observatory, 2-21-1 Osawa, Mitaka, Tokyo
181-8588,Japan\\
}
\begin{document}

%\date{Accepted 2003 December 15. Received 2003 May 03; in original form
%2003 May 03}

\pagerange{\pageref{firstpage}--\pageref{lastpage}} \pubyear{2003}

\maketitle

\label{firstpage}

\begin{abstract}
% We have investigated redshift evolution of the
% morphology--cluster-centric-radius relation and
% colour--cluster-centric-radius relation
We have investigated the redshift evolution of the
 colour--cluster-centric-radius relation and the 
morphology--cluster-centric-radius relation
%%%%%%%%%%%%%%%%%%%%%%%%%%%%%%%%%%%%%%%%%%
 in three redshift bins,
  $0.02\leq z\leq 0.14$, $0.14< z\leq 0.20$, and  $0.20<
 z\leq 0.30$
%%%%%%%%%%%%%%%%%%%%%%%%%%%%%%%%%%%%%%%%%%
 using a  homogeneous sample of 736 %328
 galaxy clusters selected from the Sloan
 Digital Sky Survey.
%%%%%%%%%%%%%%%%%%%%%%%%%%%%%%%%%%%%
 Both the relations are well-defined in all
 the redshift bins; the fraction of blue/late-type galaxies
 increases toward outside of clusters.
%%%%%%%%%%%%%%%%%%%%%%%%%%%%%%%%%%%%
%It is found that the blue/late-type galaxy
% fractions decreases with decreasing redshift at any cluster-centric
Blue/late-type galaxy fractions are found to decrease
with decreasing redshift at any cluster-centric
 radius. The trend is consistent
 with the Butcher-Oemler effect and the morphological Butcher-Oemler
% effect. In addition, we find that spectral (colour) evolution is almost
effect. In addition, we find that colour (spectral) evolution is almost
% completed at the intermediate redshift, whereas morphological evolution
 completed by $z\sim0.2$, whereas morphological evolution
%continues to the lowest redshift. The slope of the colour-radius
%relation is smooth for
%all the redshift bins, while the slope of the morphology--radius
%relation has a break only at the highest redshift. It is also found that
%fractions of blue-late type galaxies mainly decreases between  the
%highest and the intermediate redshift bins, whereas fractions of
%red-late type galaxies continuously decreases with decreasing
%redshift.
 %mainly decreases between the intermediate and the
 %lowest redshift bins.
continues to the present day. The colour-radius
relation is smooth in
all the redshift bins, while the morphology--radius
relation has a break only in the highest redshift bin.
It is also found that
%fractions of blue-late type galaxies mainly decreases between  the
%highest and the intermediate redshift bins, whereas fractions of
%red-late type galaxies continuously decreases with decreasing
%redshift.
fractions of blue-late type galaxies decreases mostly between  the
highest and the intermediate redshift bins, while fractions of
red-late type galaxies continuously decreases with decreasing
redshift through all the redshift bins.
 These results are consistent with the interpretation that
%the timescale of the spectral (colour) evolution is shorter than
the timescale of the colour (spectral) evolution is shorter than
 that of the morphological evolution.
%red-late type galaxies in the middle of the transformation
% are likely to be observed as passive spiral galaxies.
It is suggested that red-late type galaxies in the middle of
the transformation are observed as passive spiral galaxies.

% suggesting that there might exist
% unevolved red, late-type galaxies at $R_{0.7Mpc}\sim 0.3$ at the
% highest redshift.

\end{abstract}


\section{Introduction}\label{intro}

 It is a remarkable feature of galaxy population that there exists a
 correlation between galaxy type-mix and environment.
 The correlation between morphology and environment
 has been studied by many authors (Dressler 1980; Postman \& Geller 1984;
 Whitmore et al. 1993; Whitmore 1995; Dressler et
 al. 1997;  Hashimoto \& Oemler 1999; Fasano et al. 2000;  Tran et
 al. 2001; Dom{\'{\i}}nguez et al. 2001, 2002; Helsdon \& Ponman
 2003; Treu et al. 2003; Goto et al. 2003e).
% About spectral types (colours), it is known that the fractions
% of blue galaxies increases in the outer parts of clusters (e.g.,
% Butcher \& Oemler 1984; Rakos et al. 1997; Margoniner et al. 2000).
There is also a correlation between colours (spectra) and
environment. It is known that the fractions
of blue galaxies increases toward outer parts of clusters (e.g.,
 Butcher \& Oemler 1984;  Rakos \& Schombert 1995; Margoniner \& de
 Carvalho 2000; but see Andreon et al. 2003).
%%%%%%%%%%%%%%%%%%%%%%%
% ===> move to a later paragraph
% It is also worth noting that E+A (k+a or post-starburst) galaxies,
% which often thought to be cluster-related,
% are found to have their origin in merger/interaction with accompanying
% galaxies (Goto et al. 2003c,d), and thus E+A galaxies are not likely to be
% a product of the morphological transition in cluster regions.
%%%%%%%%%%%%%%%%%%%%%%%
% If we understand what these correlations imply,
% it could be a big clue to understand the origin of the types.

 Various physical mechanisms have been proposed to explain the
 correlations. Possible mechanisms include ram-pressure stripping
 of gas (Gunn \& Gott 1972; Farouki
 \& Shapiro 1980; Kent 1981; Fujita \& Nagashima 1999;
  Abadi, Moore \& Bower 1999; Quilis, Moore \& Bower
 2000);
  galaxy harassment via high speed impulsive
  encounters (Moore et al. 1996, 1999; Fujita 1998); cluster
 tidal forces (Byrd \& Valtonen 1990; Valluri 1993; Fujita 1998; Gnedin 2003a,b) which
distort
 galaxies as they come close to the centre; interaction/merging of
 galaxies (Icke 1985; Lavery \& Henry 1988; Mamon 1992; Makino \& Hut
  1997; Bekki 1998;  Finoguenov et al. 2003a);  evaporation of the cold
 gas in disc galaxies via heat conduction from the surrounding hot ICM
 (Cowie \& Songaila 1977; Fujita 2003); and a gradual decline in
 the SFR of a galaxy due to the stripping of halo gas (strangulation or
 suffocation; Larson, Tinsley \& Caldwell 1980; Bekki et al. 2002;
 Kodama et al. 2001; Finoguenov et al. 2003b).
%Although these processes are all plausible, unfortunately, there exists
%little evidence demonstrating that any one
%of these processes is actually responsible for driving galaxy
%evolution.
%Most of these processes act over an extended period of
%time, while observations at a certain redshift cannot easily provide
%the detailed information that is needed to elucidate subtle and
%complicated cluster-related processes.
Although these processes are all plausible, however, there exists
little evidence demonstrating that any one
of these processes is actually responsible for driving galaxy
evolution.
Since most of these processes act over an extended period of
time, observations at a certain redshift cannot easily provide
the detailed information that is needed to elucidate subtle and
complicated cluster-related processes.
%%%%%%%%%%%%%%%%%%%%%%
It is worth noting that E+A (k+a or post-starburst) galaxies,
which often thought to be cluster-related,
are found to have their origin in merger/interaction with accompanying
galaxies (Goto et al. 2003c,d), %; Poggianti et al. 2003)
 and thus
E+A galaxies are not likely to be
a product of the morphological transition in cluster regions.
%%%%%%%%%%%%%%%%%%%%%%%

 To shed light on the underlying physical mechanisms,
% we can trace the redshift evolution of
 it is important to trace the redshift evolution of
 correlations among morphology, colour and environment.
%  Studies on the evolution of morphology--density relation
% were attempted by Dressler et al. (1997), who reported the significant
% decrease in S0/E ratio toward increasing redshift to $z\sim 0.4$.
% The evolution of morphology--radius relation is presented by
% Fasano et al. (2000).
  Studies on the evolution of morphology--density relation
were attempted by Dressler et al. (1997) and Fasano et al. (2000),
who reported the significant decrease in S0/E ratio toward
increasing redshift up to $z\sim 0.4$.
However, such studies were hampered by the inherent difficulty
due to small and inhomogeneous cluster samples (e.g., Andreon 1998).
They compared many nearby {\it poor} clusters observed
with ground-based telescopes with the 10 {\it rich} clusters
at high redshift observed with the Hubble Space Telescopes.
%However, it has been difficult to compare the details
% (slope and absolute fractions) of such type--environment relations
% between high and low redshift due to small \& inhomogeneous cluster
%samples.
% In fact, previous study compared many nearby poor clusters observed
% with ground-based telescopes with the  10 rich clusters
% at high redshift observed with the Hubble Space Telescopes.
%About the spectral evolution of cluster galaxies,
 On the colour evolution of cluster galaxies,
 it has been known that the fractions of blue
 galaxies in a cluster increases toward higher redshift (the
 Butcher--Oemler effect; Butcher \& Oemler 1978,1984;
 Rakos \& Schombert 1995; Margoniner \& de Carvalho 2000; Margoniner et al. 2001;
 Goto et al. 2003a).
 To extract more information,
%it is important to clarify the mutual relationship between the
%spectral (colour)  evolution and the morphological evolution, in
%addition to their radial dependence.
it is important to investigate the mutual relationship between the
correlations among colour, morphology, and environments.

 With the advent of the Sloan Digital Sky Survey (SDSS; Stoughton et
% al. 2002), we now have a chance to overcome these limitations. One of
 al. 2002), we now have a chance to step forward. One of
 the  largest cluster catalogs with a well defined selection function is
 compiled from the SDSS (Goto et al. 2002a,b).
%In addition to the availability of the five optical colours
%in the SDSS ($u,g,r,i,$ and
%$z$; Fukugita et al. 1996),  since both high and low
%redshift  clusters are detected using the same cluster finding
% algorithm (Goto et al. 2002a), the
% redshift related systematic bias is minimum for this catalog.
 In addition to the availability of the five optical colours
 in the SDSS ($u,g,r,i,$ and $z$; Fukugita et al. 1996),
 this catalog has an advantage in that 
%that the redshift related systematic
%bias is minimum since 
 both high and low redshift clusters are detected using the same cluster finding
 algorithm (Goto et al. 2002a).
 In this study, we adopt the cluster-centric-radius as a parameter of
 environment, %rather than local density
 and investigate the evolution of colour--radius relation
 and morphology--radius  relation.

 The paper is organized as follows: In
 Section \ref{data}, we describe the SDSS data we used.
 In Section \ref{analysis}, we explain automated morphological
 classifications and the method to estimate type fraction.
%% density estimation.
 In Section \ref{results}, we present the results and discuss the
 physical implications.
% Section \ref{conclusion}, we summarize our work and findings.
   The cosmological parameters adopted throughout this paper are $H_0$=75 km
 s$^{-1}$ Mpc$^{-1}$, and
($\Omega_m$,$\Omega_{\Lambda}$,$\Omega_k$)=(0.3,0.7,0.0).

\section{The SDSS Data}\label{data}
 
 The data used in this study are essentially the same data as used in
 Goto et al. (2003a). The galaxy catalog is taken from  the Sloan
 Digital Sky Survey Early Data Release (SDSS EDR; 
 Stoughton et al. 2002), which covers $\sim$394 deg$^2$ of the sky  
 (see Fukugita et al. 1996; Gunn et al. 1998;  Lupton
  et al. 1999,2001,2002; York et al. 2000; Eisenstein et al. 2001;
  Hogg et al. 2001; Blanton et al. 2003; Richards et al. 2002;
 Stoughton et al. 2002; Strauss et   al. 2002; Smith et al. 2002;  Pier et
  al. 2003  for more detail  of the SDSS data). The cluster catalog is
 presented in Goto et al. (2002a).  In this study, we limit our
 sample clusters to those with richness between 20 and 80, using the cluster
 richness measured as a number of galaxies between $M_{r^*}=-24.0$
 and $-19.44$ within 0.7 Mpc from a cluster centre after
 fore/background subtraction (see Goto et al. 2002b).
%% We used $k$-correction given in Fukugita et al. (1995).  
 These richness cuts nicely select rich clusters that allows us to
 study radial profile, and at the same time to reject richest clusters
 minimizing a possible richness related bias.
 The resulting number of clusters in our sample is still large ($N=736$).
 We separate these clusters in three
 redshift bins; $0.02\leq z\leq 0.14$, $0.14< z\leq 0.20$, and  $0.20<
 z\leq 0.30$ (corresponding lookback time of 0.9, 1.9 and 2.7 Gyr). Each
 redshift bin has 82, 188 and 466 clusters, respectively.  

\section{Analysis}\label{analysis}

%\subsection{Galaxy Classification}\label{morphology}
 
\begin{figure*}
  %\vspace*{174pt}
\vspace*{20pt}
\includegraphics[scale=0.4]{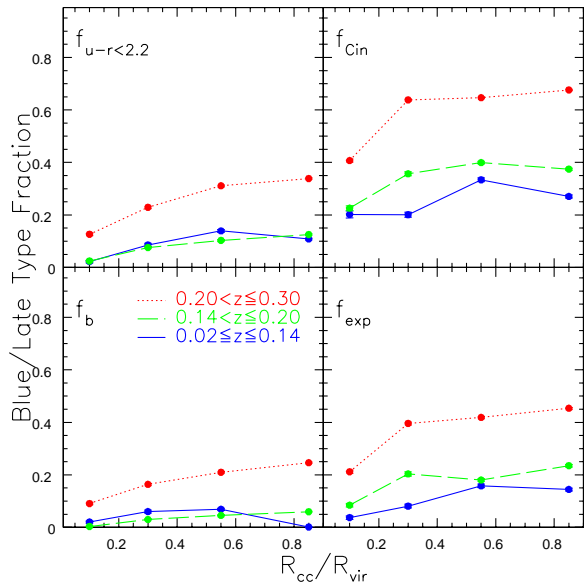}
\caption{
 Evolution of the morphology/colour--radius relation. The
 fractions of blue/late-type galaxies are plotted against
 cluster-centric-radius normalized using cluster richness. The solid,
 dashed and dotted lines represent clusters with $0.02\leq z\leq 0.14$,
 $0.14< z\leq 0.20$, and  $0.20< z\leq 0.30$.
}\label{fig:mr} 
\end{figure*}

 In order to measure blue/late-type fractions,
 we have to subtract fore/background galaxy counts statistically 
 since our galaxy catalog is based on the imaging data.  
 The adopted procedure is the same to the one used in Goto et al. (2003a)
 except the use of global galaxy number counts. 
 Instead of local galaxy number counts, we use the entire 394 deg$^2$ of
 the imaging data to calculate an 
 average galaxy number counts in the 
 magnitude range between $M_{r^*}=-24.0$ and $-20.7$. 
 The use of global fore/background estimation is essential to probe
 slight over-densities in cluster perimeter regions. 
 Among the entire
 region, cluster regions are only 2.4\% in angular area.
 We re-scale this average galaxy number counts to meet an angular area of
 each radial bin, then subtract fore/background galaxy counts
 statistically.

 Since a physical size of a cluster differs cluster by cluster, it is important
 to normalize the cluster radius when measuring blue/late type
 fractions. Since it is difficult to measure  a virial radius using
 the imaging data, we measure a pseudo-virial radius in a following way to
 normalize the cluster-centric radius (see Goto et al. 2003a for details).
%It is
% ideal to use a virial radius to normalize the cluster radius. However,
% since such a measurement is not available to our sample clusters taken
% from the  imaging data, we use cluster richness to normalize the radius as
% follows (see Goto et al. 2003a for details).
\begin{equation} 
    R_{vir} = 0.7\times(Richness/32)^{1/3} (Mpc)\label{eq:radius}
\end{equation}
 %Since the median richness of the cluster sample is 32,
  Equation (\ref{eq:radius}) returns us with a
 normalized cluster radius of $R_{vir}=0.7$ Mpc for clusters with
 $Richness=32$ (c.f., 0.7 Mpc $\simeq$ a virial radius of a system with
 $\sigma =$ 350 km s$^{-1}$). We note that $R_{vir}$ is a
 pseudo-virial radius, and that caution is needed when comparing with
 a virial radius from a more realistic measurement such as velocity
 dispersion or X-ray radial profile.
 Using a position of the brightest cluster galaxy given in Goto et
 al. (2002a) as a cluster centre, we sum the number of blue/late
 and all galaxies within radial bins separated at $R_{cc}$/$R_{vir}=0.2, 0.4,
 0.7$ and $1.0$ ($R_{cc}$ is a distance from a cluster centre.), and
 then subtract fore/background statistically to  
 measure blue/late-type fractions.

 To classify galaxies into blue/late-type and others, we use exactly the 
 same method as in Goto et al. (2003a), which we briefly summarize
 below. We use galaxies between  $M_{r^*}=-24.0$ and
 $-20.7$. This absolute magnitude range assures the proper
 measurement of 
 $u-r$ and $g-r$ colours for the faintest galaxies at the highest
 redshift bin.   
 Fractions of blue galaxies are defined in two different ways;  $f_b$
 and $f_{u-r<2.2}$. 
 The first method, $f_b$, uses restframe $g-r$ colour. If a galaxy is
 bluer by more than 
 0.2 in $g-r$ than the colour of the red-sequence of elliptical galaxies
 at the redshift of the cluster, we define the galaxy as blue. This
 definition is a similar definition as was used in original Butcher \&
 Oemler (1984) and recently in Margoniner et al. (2001). 
 The fraction
 of blue galaxy, $f_b$, is defined as the ratio of a number of blue
 galaxies to that of all galaxies after the global fore/background subtraction. 
 The second method, $f_{u-r<2.2}$, defines a blue galaxy as
 a galaxy with 
 $u-r<2.2$ (Strateva et al. 2001). Again, the fraction
 of blue galaxy,  $f_{u-r<2.2}$, is defined as the ratio of a number of
 blue galaxies to that of all galaxies after the global fore/background subtraction.

 To separate galaxies into early/late-type morphologically, we use two
 different purely morphological parameters as are used in Goto et
 al. (2003a).  The first one, $f_{Cin}$, uses a concentration parameter, $Cin$,
 defined as the ratio of Petrosian 50\% flux radius to  Petrosian 90\%
 flux radius. We regard a galaxy with $Cin>0.4$ as late-type
 (Shimasaku et al. 2001).  Note that  $f_{Cin}$ can be affected
 by the seeing size. Goto et al. (2003a) estimated that an increasing
 seeing size might increase late-type fractions by $\sim$5\% between
 $z=0.3$ and $z=0.02$.  The other morphological classification, $f_{exp}$, uses
 galaxy radial 
 profile fitting. When an exponential profile fits better than de
 Vaucouleurs' profile to a galaxy, the galaxy is classified into
 late-type.  The fraction
 of late-type galaxy ($f_{Cin}$ or $f_{exp}$) is defined as the ratio of
 a number of 
 late-type galaxies to that of all galaxies after the global
 fore/background subtraction.

 The errors on blue/late-type fractions are computed 
 based on equation (2) of Goto et al. (2003a). 
 The errors are as small as the dot size in Figure \ref{fig:mr}. 
 Note that these errors do not consider errors in classifying galaxies
 into sub-categories. 
 
\section{Results \& Discussion}\label{results}

\begin{figure*}
  %\vspace*{174pt}
\vspace*{20pt}
\includegraphics[scale=0.4]{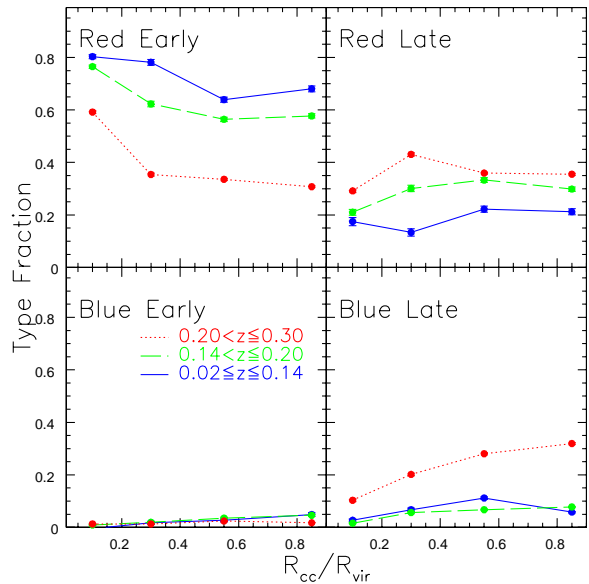}
\caption{
 The fractions of four types of galaxies are plotted against
 cluster-centric-radius normalized using cluster richness. The criterion
 to separate red/blue is $u-r=2.2$, and that for early/late is $Cin=0.4$.
 The solid,
 dashed and dotted lines represent clusters with $0.02\leq z\leq 0.14$,
 $0.14< z\leq 0.20$, and  $0.20< z\leq 0.30$.
}\label{fig:mr_yagi} 
\end{figure*}

 We present the evolution of 
 the colour--radius relation
 and 
the morphology--radius relation
 based on 
 the SDSS data in Figure \ref{fig:mr}, where blue/late type fractions
 are plotted against the normalized cluster-centric-radius for the three redshift
 bins. We use four annulus bins for radial direction (each separated at
 $R_{cc}$/$R_{vir}=0.2, 0.4, 0.7$ and $1.0$) and three redshift
 bins ($0.02\leq z\leq 0.14$, $0.14< z\leq 0.20$, and  $0.20< z\leq
 0.30$). The $f_b, f_{u-r<2.2}, f_{Cin},$ and $f_{exp}$ are plotted in the
 lower left, upper left, upper right, and lower right panels,
 respectively; i.e., the two left panels probe colour evolution of
 cluster galaxies and the two right panels are for morphological
 evolution. 
 The following observational results can be recognized from Figure
 \ref{fig:mr}.  

  In all the panels, the fractions of blue/late type galaxies decrease
  with decreasing redshift at almost all cluster-centric-radius. The decrease
  is consistent with the 
  Butcher--Oemler effect (left panels) and the morphological
  Butcher--Oemler effect (right panels).

 The amount of decrease in fractions might be smaller for colour
 evolution (left panels; e.g., less than 20\% decrease between the
 highest and the 
 lowest redshift bins) than for morphological evolution (right panels;
 greater than 30\% decrease between the highest and the
 lowest redshift bins). If the trend is real, it might
 imply that some red, late-type galaxies changed their
 morphology into early-type between $z=0.3$ and $z=0.02$.
% However, the
% morphological parameters in the right panels are more easily affected
% by the seeing size. Goto et al. (2003a) estimated that an increasing
% seeing size might increase late-type  fractions by $\sim$5\% between
% $z=0.3$ and $z=0.02$. Therefore, the trend should not be
% over-interpreted. 

 Furthermore, at the intermediate redshift ($0.14< z\leq 0.20$), the
 fractions of 
 blue galaxies in the two left panels are almost as low as those at the
 lowest redshift bin ($0.02\leq z\leq 0.14$). On the other hand,
 morphological late-type fractions shown in the two right panels are
 $\sim$10\% larger than those at  the lowest redshift bin. The trend
 might suggest that the colour (spectral) evolution is already completed at the
 intermediate redshift bin, while the morphological evolution continues to
 the lowest redshift bin; i.e., timescale of the colour (spectral) 
 evolution may be shorter than the timescale of the morphological evolution. 
% The trend is
% consistent with the discovery of passive spiral galaxies (Couch et
% al. 1998; Poggianti et al. 1999; Goto et al. 2003b), which
% have red colours with spiral morphology, also suggesting that the spectral
% evolution is quicker than the morphological evolution.

 The radial profile are smooth for the fractions of blue galaxies
 at any redshift bin. However, for late-type fractions in the two right
 panels, the radial profiles seem to have a break at $R_{cc}$/$R_{vir}
 \sim 0.3$ at the highest redshift bin. Then, the profiles
 become smoother 
 with decreasing redshift. %, showing profiles as smooth as those of blue
 %fractions at the lowest redshift bin.  
 A similar trend is found by Treu
 et al. (see their Fig. 17) in comparing the morphology--density relation
 of their cluster Cl0024$+$16 with that of the nearby cluster sample in
 Dressler (1980). 
 The result indicates that the
 delayed morphological evolution occurs
 around  $R_{cc}$/$R_{vir}=$ 0.3--0.8, where many spectrally
 evolved, but morphologically unevolved galaxies possibly exist.  The
 interpretation is consistent with Goto et al. (2003b), who found
 passive spiral 
 galaxies in the infalling region (perimeter) of clusters. Note that
 Goto et al. (2003b) used galaxies with spectroscopy
 whereas this work statistically determines galaxy fractions from the
 imaging data (see also Couch et al. 1998; Poggianti et al. 1999 for
 passive spirals). 

 The difference in timescale of the spectral evolution and the
 morphological evolution is further clarified in Figure
 \ref{fig:mr_yagi}, where we plot fractions of four different types of
 galaxies as a function of cluster-centric-radius. Here, we classified
 galaxies into the following four different subsamples: 
 blue-late  ($u-r<2.2$, $Cin\geq 0.4$), 
 blue-early ($u-r<2.2$, $Cin\geq 0.4$), 
 red-late   ($u-r\geq 2.2$, $Cin<0.4$), and
 red-early  ($u-r\geq 2.2$, $Cin<0.4$).
 In the lower right and upper left panels, we plot the redshift
 evolution of blue-late and 
 red-early types of galaxies as a function of cluster-centric-radius. As
 expected, red-early type shows gradual increase with decreasing
 redshift and blue-late type shows gradual decrease with decreasing
 redshift. However, a contrasting result is found when we compare
 red-late fractions (upper right) with blue-late fractions (lower right).  The
 fractions of red-late  
 type are not very much dependent on the cluster-centric-radius. In
 addition, red-late fractions show much evolution between the lowest and
 the intermediate redshift bins, whereas blue-late fractions show large
 evolution only between the highest and intermediate redshift bins.
 Assuming
 that blue galaxies evolve into red, and that late type evolves into early
 type,  these results suggest that colour evolution (blue to red)
 mainly occurs between the highest and intermediate redshift, and that
 morphological evolution (late to early) continues toward the lowest
 redshift even after colour evolution was completed.
% A contrasting result can be found in the fraction of red-late
% type of galaxies in the upper left panel. The fractions of red-late
% type are not very much dependent on the cluster-centric-radius. In
% addition, compared with the results at the highest redshift (solid
% line) the fractions of red-late types do not show much difference  at
% the intermediate redshift (dashed line), and then decreases at
% the lowest redshift 
% (dotted line). This result is consistent with the interpretation that
% between the highest and intermediate redshift, mainly spectral evolution
% takes place, creating spectrally evolved, but morphologically unevolved
% red-late type galaxies; then, between the intermediate and the lowest
% redshift, morphological evolution catches up with the spectral
% evolution, reducing the fractions of red-late type galaxies.
 In other
 words, it might indicate  that morphological evolution needs longer
 timescale to be 
 completed ($\sim$2 Gyr) than the colour evolution does ($\sim$1 Gyr). 
%The result is consistent 
% with our findings in Figure \ref{fig:mr}. 
 These estimates will provide  a good benchmark to
 compare with semi-analytic simulations (e.g.,  Okamoto \& Nagashima
 2001; Diaferio et al. 2001; Benson et al. 2001; Springel et al. 2001;  Okamoto \& Nagashima 2003).
   Finally, we would like to mention that  red-late type galaxies may be the same galaxy population as passive spiral galaxies found in Goto et al. (2003b).

%\section{Conclusions}\label{conclusion}

\section*{Acknowledgments}

% We are grateful to XXX for useful discussions.
% We are grateful to Spiderman for keeping the world at peace.
% We wish to express our gratitude to XXX for useful conversation.  
 We thank the anonymous referee for many insightful comments which
 improved the manuscript significantly. 
 T. G. acknowledges financial support from the Japan Society for the
 Promotion of Science (JSPS) through JSPS Research Fellowships for Young Scientists.
%
% The Sloan Digital Sky Survey (SDSS) is a joint project of The
%University of Chicago, Fermilab, the Institute for Advanced Study, the
%Japan Participation Group, the Johns Hopkins University, the
%Max-Planck-Institute for Astronomy, New Mexico State University,
%Princeton University, the United States Naval Observatory, and the
%University of Washington. Apache Point Observatory, site of the SDSS
%telescopes, is operated by the Astrophysical Research Consortium
%(ARC).  Funding for the project has been provided by the Alfred
%aP.~Sloan Foundation, the SDSS member institutions, the National
%Aeronautics and Space Administration, the National Science Foundation,
%the U.~S.~Department of Energy, Monbusho, and the Max Planck
%Society. The SDSS Web site is http://www.sdss.org/.
%

%\bsp

\label{lastpage}

\end{document}